\documentclass[10pt]{article}
\pdfoutput=1

\usepackage{cite}

\usepackage[nointlimits,reqno]{amsmath}
\usepackage{amssymb}
\usepackage{amsmath}
\usepackage{graphicx}
\usepackage{tabularx}
\usepackage{multirow}
\usepackage{epstopdf}
\usepackage{slashed}
\usepackage{verbatim}
\usepackage[section]{placeins}
\usepackage[space]{grffile}
\usepackage{placeins}
\usepackage{feynmf}
\usepackage{float}

\numberwithin{equation}{section} % Equation numbering within sections

\usepackage[pdftex,
 pdfproducer=TeXShop,
 pdfcreator=pdflatex]{hyperref}

\usepackage{xspace}
\usepackage{luty}

\newcommand{\ohw}{$\mathcal{O}_{HW}\hspace{3pt}$}
\newcommand{\ohb}{$\mathcal{O}_{HB}\hspace{3pt}$}

\begin{document}

% =============================================================================
% Title page
% =============================================================================
\begin{titlepage}
%\preprint{}

\title{Dimension-6 Operator Constraints from Boosted VBF Higgs}

\author{Ralph Edezhath }
%\footnote{ \href{mailto:raedezhath@ucdavis.edu} {\tt raedezhath@ucdavis.edu } }
\address{Physics Department, University of California,\\

Davis, California 95616}

\begin{abstract}
We discuss the constraints on new physics from Higgs production through vector boson fusion in the context of an effective field theory that preserves Standard Model gauge symmetries. We find that the constraints on dimension-6 operators are significantly improved over those from the VBF signal strength by studying the  Higgs transverse momentum distribution.  Focusing on the $\mathcal{O}_{HW}$ operator, we find that boosted VBF decaying to photons yields constraints competitive with boosted WW production in the fully leptonic final state, and calculate projected limits for both at the 14 TeV LHC. The $p_T$ cuts required to maximize the the reach of VBF searches are substantially softer, making the use of the effective field theory more robust than in the case of WW production which requires very high $p_T$ cuts to obtain similar limits. Boosted VBF Higgs is thus an important probe of new physics.

\end{abstract}

\end{titlepage}

% =============================================================================
% =============================================================================
\section{Introduction}
\label{sec:intro}
Although the existence of the Higgs boson has been firmly established \cite{HiggsATLAS,HiggsCMS}, the naturalness argument indicates that our understanding of electroweak symmetry breaking may be incomplete. Furthermore, the absence of any deviations of Higgs properties from Standard Model (SM) expectations suggests that new physics, if it exists, may be decoupled at some heavy scale. This motivates a comprehensive program of precision measurements of the Higgs interactions to detect hints of new physics. At energies below the heavy scale of new states, they can be integrated out giving rise to an effective field theory (EFT) of higher dimensional operators composed of SM fields. The EFT for new physics above the weak scale which preserves the $SU(3)_C\times SU(2)_W\times U(1)_Y$ symmetry of the SM was first formulated in\cite{dim6}. This EFT provides a model independent framework for interpreting precision measurements, which can be connected to specific UV models systematically; a recent discussion can be found in \cite{EFTmatching}.  

Constraints on these operators have been derived from electroweak precision measurements \cite{ewpt1,ewpt2,Falkowski:2014tna}, from triple gauge couplings (TGC) \cite{tgc1,tgc2} and various Higgs sector measurements\cite{HiggswindowsPomarol,Pomarol:2013TowardstheUltimateSMFit,Falkowski:2013dza,offshell}. Global fits incorporating various searches have been performed using electroweak and TGC data\cite{Skiba EWPTfit} and later including Higgs sector constraints\cite{Corbett:2012RobustDeterminationoftheHiggsCouplings,Ellis:2014CompleteHiggsSectorConstraintsonDimension6,Ellis:2014TheEffectiveStandardModelafterLHC,deBlas:2014ula}. 
Projected constraints from future lepton colliders were studied in \cite{HiggstralungILC,Beneke:2014vqa}. The constraints from the vector boson fusion (VBF) production of Higgs have been relatively unexplored. The use of angular correlations in VBF production to probe the spin and CP of the Higgs was studied in \cite{vbfangular}. The boosted signatures of dimension-6 operators were studied for gluon fusion Higgs\cite{boostedggf,Ghosh:2014wxa} and Higgs plus vector boson production\cite{Ellis:2014CompleteHiggsSectorConstraintsonDimension6,Biekoetter:2014jwa} but the constraints from boosted VBF have not been examined. The constraints on these operators from the signal strength ratio of VBF to gluon fusion $\frac{\mu_{VBF}}{\mu_{ggF}}$ was studied in \cite{Ellis:2014CompleteHiggsSectorConstraintsonDimension6}, combining the signal strength likelihoods reported by CMS and ATLAS for $\gamma\gamma,\bar{\tau}\tau,ZZ^*$ and $WW^*$. We find that a stronger constraint can be obtained from boosted VBF than from the signal strength alone.
  
Many bases have been proposed for these operators\cite{dim62, Hagiwarabasis,Grojeanbasis}; we use the `Strongly Interacting Light Higgs' (SILH) basis which was first proposed in \cite{SILH} and extended in \cite{SILH2}. The bosonic operators which modify VBF in this basis are $\mathcal{O}_H$,$\mathcal{O}_W$, $\mathcal{O}_B$, $\mathcal{O}_{BB}$, \ohw and \ohb which are respectively:
\def\lra#1{\overset{\text{\scriptsize$\leftrightarrow$}}{#1}}
\begin{align*}
 & \Delta \mathcal{L}_{VBF} =  \dfrac{c_H}{2\Lambda^2} (\partial^\mu |H|^2)^2+  \dfrac{ig' c_B}{2 \Lambda^2}\left( H^\dagger  \lra {D^\mu} H \right )\partial^\nu  B_{\mu \nu}+ \dfrac{ c_{W}}{2\Lambda^2}\left( H^\dagger  \sigma^a \lra {D^\mu} H \right )D^\nu  W_{\mu \nu}^a \\ & + \dfrac{ c_{BB} {g}^{\prime 2}}{\Lambda^2} |H|^2 B_{\mu\nu}B^{\mu\nu}  + \dfrac{ ig c_{HW}}{\Lambda^2}   (D^\mu H)^\dagger\sigma^a(D^\nu H)W^a_{\mu\nu}+ \dfrac{ig' c_{HB} }{\Lambda^2}(D^\mu H)^\dagger(D^\nu H)B_{\mu\nu}  .
\end{align*}
\\ 
The operators $\mathcal{O}_W$ and $\mathcal{O}_B$ are tightly constrained by the S parameter while $\mathcal{O}_{BB}$ is constrained by Higgs to diphoton decay \cite{Pomarol:2013TowardstheUltimateSMFit}, and $\mathcal{O}_H$ modifies the Higgs propagator. The $\mathcal{O}_{HW}-\mathcal{O}_{HB}$ direction is constrained by the decay of Higgs to Z$\gamma$\cite{Falkowski:2013dza}, but this vanishes if the new physics giving rise to these operators obeys $P_{LR}$ symmetry\footnote{The parity that interchanges $L \leftrightarrow R$ for $SU(2)_L\times SU(2)_R$. For a discussion of the behavior of these operators under custodial symmetry and $P_{LR}$ see \cite{HiggswindowsPomarol}}, thus it is important to probe these operators through other measurements. The $\mathcal{O}_{HW}+\mathcal{O}_{HB}$ direction contributes to anomalous triple-gauge couplings and the limits from TGC measurements at LEP and LHC have been studied in \cite{Pomarol:2013TowardstheUltimateSMFit,Ellis:2014CompleteHiggsSectorConstraintsonDimension6,Falkowski:2014tna}. The TGC limits relevant for this direction come from WW production, which is currently also the sole probe of the $\mathcal{O}_{WWW}$ operator. An independent measurement of the effect of $\mathcal{O}_{HW}+\mathcal{O}_{HB}$ is necessary to disentangle it from $\mathcal{O}_{WWW}$. VBF Higgs is thus complementary to WW production searches and one of the few probes of the direction that is allowed by $P_{LR}$ symmetry. 
 
In the following sections, we study whether the current limits from VBF Higgs are competitive with TGC limits from diboson production and calculate projected from the 14 TeV LHC for both. We restrict our analysis to the effect of \ohw, as the operator \ohb has the same behavior but suppressed by $\tan^2\theta_W$. We set $|c_{HW}|=1$ so that $\Lambda$ indicates the scale of new physics. 

\begin{figure} 
\centering \includegraphics[scale=0.45]{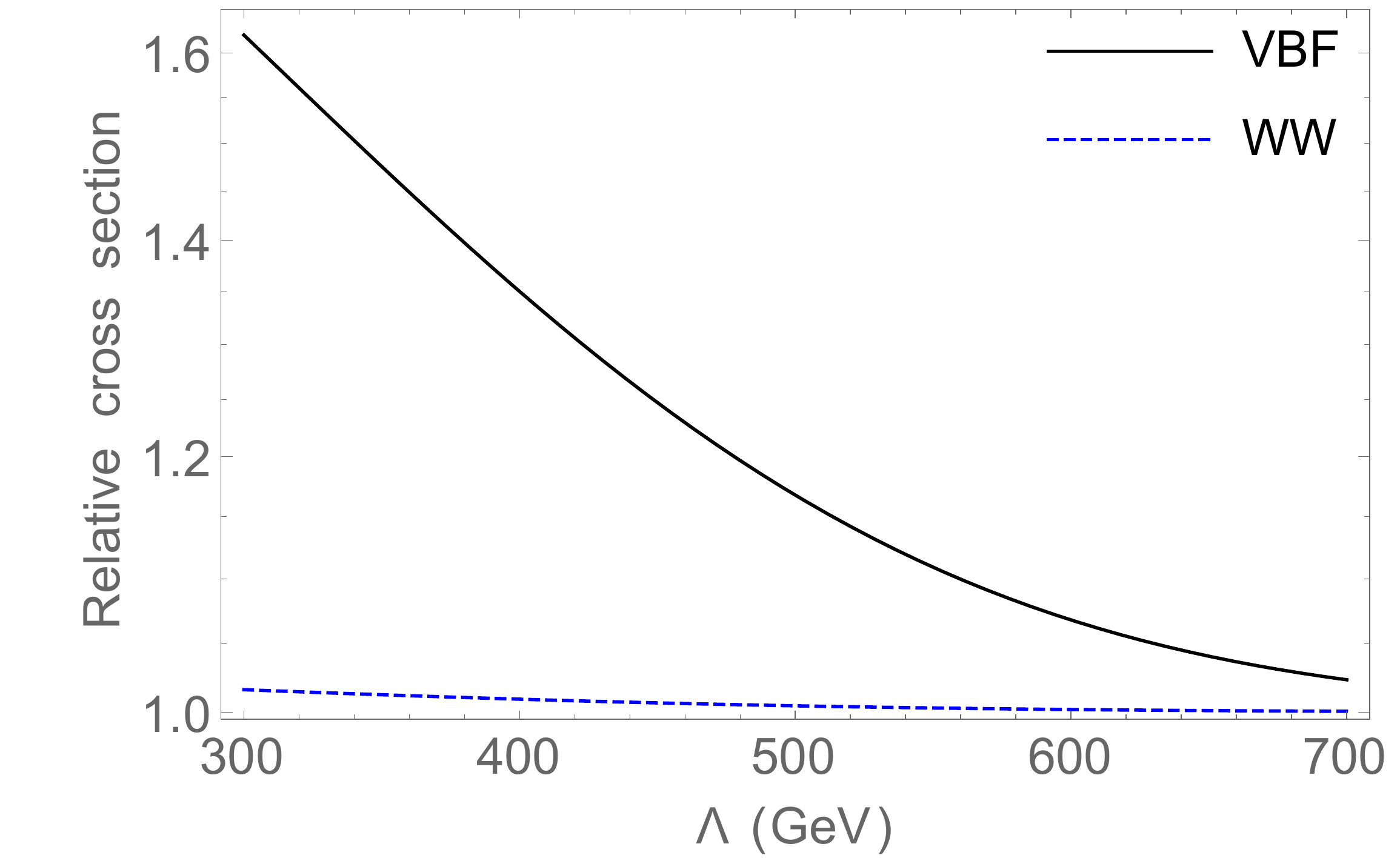}
%\begin{minipage}{5.5in}
\caption{\small 
Ratio of the cross section with \ohw over the SM cross section, for VBF and WW production  at $\sqrt{s}=14$TeV}
\label{fig:relrate}
%\end{minipage}
\end{figure}

WW production has a significantly larger rate than VBF Higgs production, and hence the uncertainty in its signal strength is much lower. However the relative enhancement due to the dim-6 operators is much higher in VBF as shown in Fig. \ref{fig:relrate}, thus it has the potential to set competitive limits. The momentum dependence in \ohw and \ohb will enhance VBF production at high Higgs $p_T$, and we examine whether this can be used to improve the limits. We will derive and compare the current and projected dim-6 operator limits for VBF Higgs decaying to photons and the fully leptonic decay of WW. We compare the CMS studies\cite{CMS:rilUpdatedmeasurementsoftheHiggsboson,CMSWW7tev} at 7 TeV since the limits for the full 8 TeV dataset have not yet been released.  The projected VBF sensitivities for $\tau\tau$, ZZ and WW are similar to the diphoton channel\cite{ATLASprojection} so we hope that the limits obtained through a study of the diphoton channel will be representative of the reach of the other channels.

\section{Limits from WW production}
So far the most stringent limits on the $\mathcal{O}_{HW}+\mathcal{O}_{HB}$ direction in LHC data have been derived from TGC measurements in WW production. The operators \ohw   and \ohb contribute to triple-gauge couplings as follows
$$
\Delta\mathcal{L}_{TGC}=i g\delta g_1^Z c_{\theta_W}
 Z^{\mu}  \left( W^{- \, \nu}  W^+_{\mu\nu} 
  - W^{+ \, \nu}   W^-_{\mu\nu} \right) 
 \nonumber 
+
ig \left(  \delta \kappa_Z c_{\theta_W} Z^{\mu\nu} + \delta\kappa_\gamma s_{\theta_W} A^{\mu\nu}\right) 
W^-_\mu  W^+_\nu
\nonumber
$$
where   $ V_{\mu\nu} \equiv  \partial_\mu V_\nu -  \partial_\nu V_\mu$ for $V=W^\pm,Z,A$ and the TGC parameters are defined as in\cite{Hagiwara:1986ProbingtheWeakBosonSector} :
\begin{align*}
& \delta g^Z_1 = c_{HW} \frac{m_z^2}{\Lambda^2}  \\ & \delta k_\gamma = (c_{HW}+c_{HB})\frac{m_w^2}{\Lambda^2} \\ &  \delta k_Z = \delta g^Z_1 - tan^2_{\theta_W} \delta k_\gamma
\end{align*}

The backgrounds for the fully leptonic decay of $WW$ are the following : $W+jets$ where one of the jets fakes a lepton, gluon induced $WW$ production, $t\bar{t}, tW,$ Drell-Yan production of leptons, $WZ, ZZ$ and $W+\gamma$ where the photon fakes an electron. The event samples for gluon-induced $WW$ production were generated using {\tt gg2VV}\cite{gg2VV}, and for the other processes the events were generated with {\tt MadGraph5 v2.1.2}\cite{MG5} interfaced with {\tt Pythia 6.4}\cite{Pythia} and {\tt Delphes 3.1.2}\cite{Delphes}, and the UFO model  implementation  of dim-6 operators from \cite{HELmodel}. We implement the following cuts from the CMS $WW$ production search in the fully leptonic final state \cite{CMSWW7tev} : two oppositely charged leptons, with $p_T$ of the dilepton system $> 45$ GeV are required, with an invariant mass of $| m_{\ell\ell} - m_Z| >15 $GeV , veto on b-jets or jets with $E_T>15$ GeV and with azimuthal angle within $165 \degree$ of the dilepton system (in the case of same-flavor leptons), no jets with $E_T>30$ GeV and $\eta<5$. The `projected MET' as defined in\cite{CMSWW7tev} is required to be greater than 37.5 (20) GeV for same flavor leptons (opposite flavor) leptons. High $p_T$ cuts on the leading lepton significantly reduces the non-$WW$ background, and since we do not consider pileup we have not made further optimizations for 14 TeV. The cross sections for the $WW$ and background processes after these cuts is given in Table \ref{tab:wwprod}.

\begin{table}[h!]

\label{tab:wwprod}
\begin{center}
\begin{tabular}{l|c|c}
Process &  Passing inclusive cuts & With lepton $p_T^{max}>760$GeV     \\
\hline 
SM $W^+W^-\rightarrow l^+\nu l^-\nu$ & 257 & 0.0017 \\
SM $W^+W^-$+\ohw with $\Lambda =$1000GeV & 257 & 0.007 \\
\hline 
$gg\rightarrow WW$ &19.9  & $8.2\cdot 10^{-4}$\\
$t\bar{t}+tW$ & 27.7 & $3.7\cdot 10^{-4}$\\
$W+jets$ & 20.5 & $1.9\cdot 10^{-4}$\\
 $Z/ \gamma^* \rightarrow l^+/ l^-$ & 1.49 & $3.0\cdot 10^{-4}$\\
$WZ$ and $ZZ$ &  3.33 & $1.4\cdot 10^{-4}$  \\
$W\gamma^*$&  6.8& $2.5\cdot 10^{-4}$ \\
\end{tabular}
\end{center}
\caption{$WW$ search cross sections (fb) at $\sqrt{s}=14$TeV. The second column shows the significant reduction in the background processes and enhancement of \ohw relative to SM-only $WW$ production from a high $p_T$ cut on the leading lepton }
\end{table}

\begin{figure} 
\centering \includegraphics[scale=0.35]{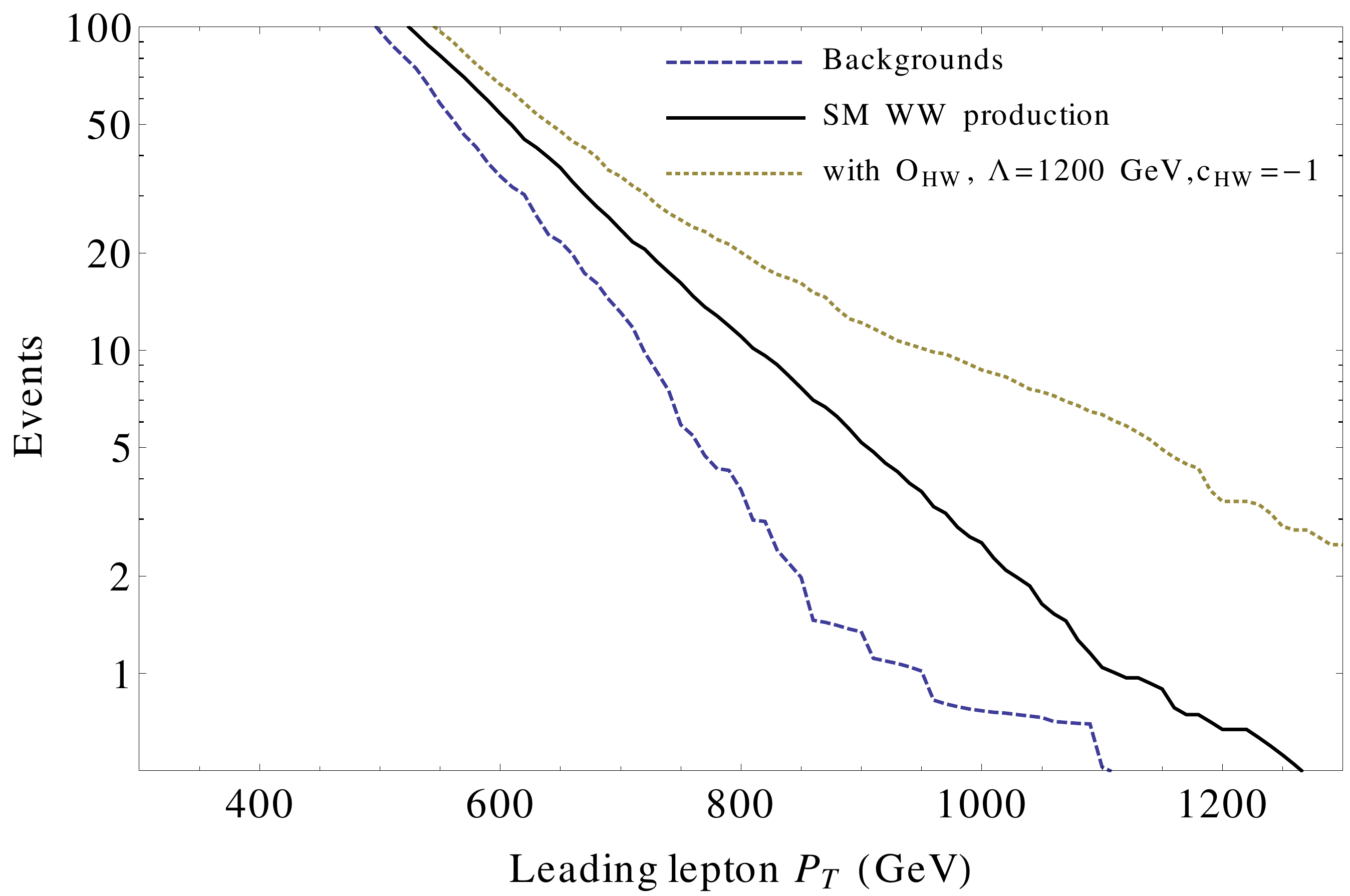}
\begin{minipage}{5.5in}
\caption{\small 
Cumulative plot for total events above given value of leading lepton $p_T$, for $\sqrt{s}=14$TeV and 300fb$^{-1}$. The sum of the backgrounds from gluon induced WW production, t$\bar{\mbox{t}}$, tW, Drell-Yan production of leptons, WZ, ZZ and W+photon are shown alongwith WW production in the SM and with \ohw}
\label{wwpt}
\end{minipage}
\end{figure}

\vspace{5mm}
The effect of the dim-6 operators are more pronounced at higher W boson $p_T$, thus the CMS search uses a maximum-likelihood fit for the $p_T$ distribution of the leading lepton to set limits on TGCs. We approximate this with a $p_T$ cut (shown in Table \ref{wwpt}) on the leading lepton corresponding to the highest bin reported by the search, since this is the most sensitive to the effect of dim-6 operators. This yields a $95 \%$ CL limit of 320 \hspace{1pt}GeV for $\Lambda$. The limit reported by the CMS search of $|\delta g_1^Z| \leq 0.095$, which corresponds to a limit on $\Lambda$ of 295 GeV. We slightly overestimate the bound since we did not consider systematics, and consider only the Poisson statistical error. The dominant systematic error arises from the difference in the $p_T$ spectrum between the $W+jets$ background and the QCD multijet spectrum, since the former is estimated from the latter. But at high $p_T$ the yield from the non-$WW$ background events is very small as shown in the Fig. \ref{wwpt}, thus the effect of this error on the dim-6 operator limits should be small. We collect the 7 TeV and projected limits below, with the corresponding $p_T$ cut required.

\begin{center}
\begin{tabular}{c|c|c|c}
\label{wwxsec}
$\sqrt{s}$ and Luminosity & $p_T$ cut on leading lepton & \multicolumn{2}{c}{95\% CL Limit on $\Lambda$}  \\
\hline
&  &  $c_{HW}=-1$  &$c_{HW}=1$ \\
\cline{3-4}
7 TeV, 4.92fb$^{-1}$ & 180 GeV & 320 GeV& 385 GeV  \\  
14 Tev, 300fb$^{-1}$ & 760 GeV &  1000 GeV&1200 GeV \\
14 Tev, 3ab$^{-1}$ & 940 GeV & 1350 GeV&1825 GeV  \\ % check hwp
\end{tabular}
\end{center}

\begin{figure}
\centering \includegraphics[scale=0.35]{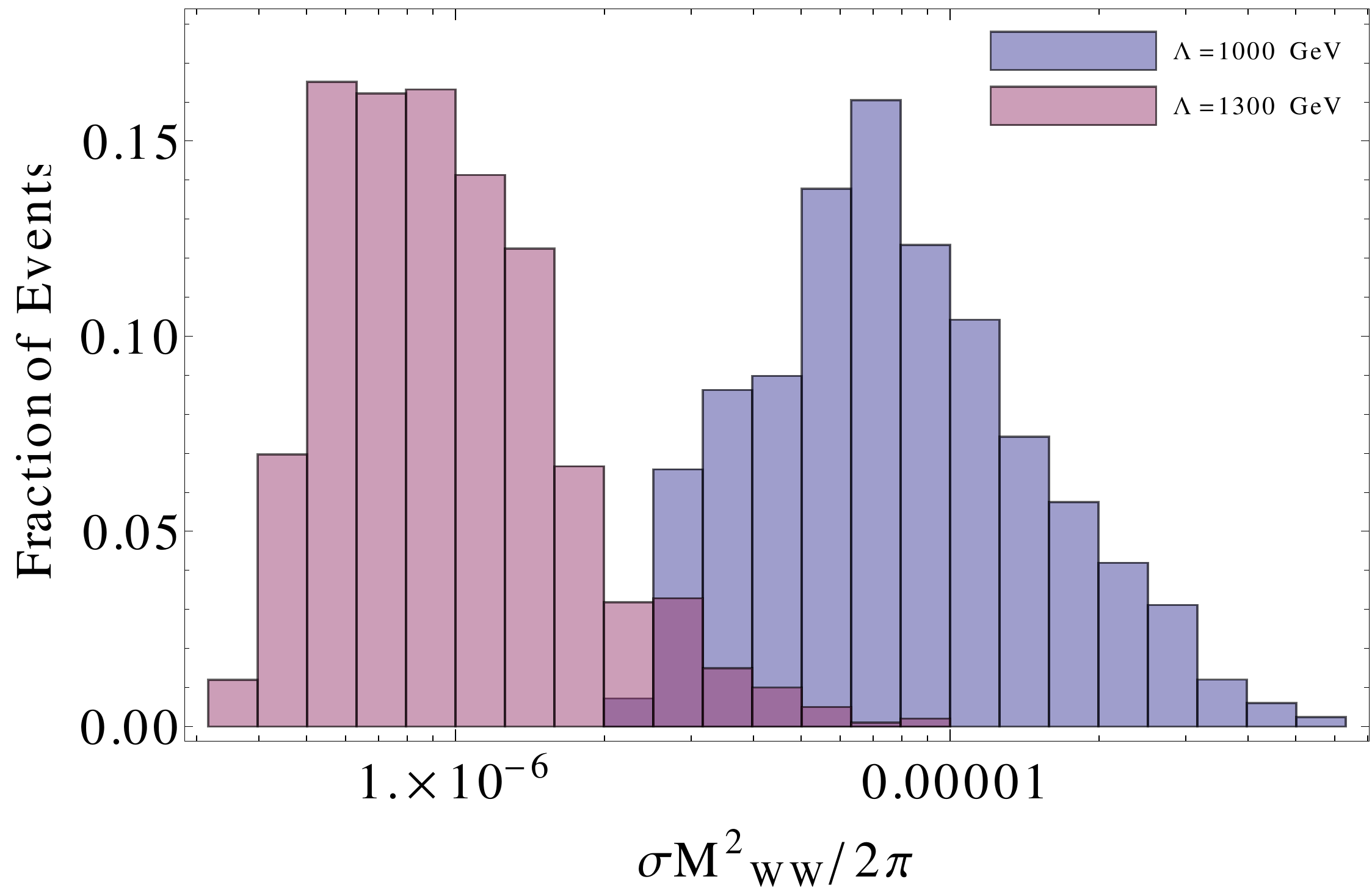}

\caption{\small 
Histogram of $\dfrac{\sigma_{tot} \cdot m^2_{WW}}{2\pi}$ for events passing the leading lepton $p_T$ cut}
\label{eftww}

\end{figure}

Since the $p_T$ cut for the leading lepton is close to the naive scale of the new physics, it is important to check whether the theory violates unitarity in the high $p_T$ region. Following\cite{EFTModernApproach}, unitarity is violated for $q\bar{q}\rightarrow WW$ when $\dfrac{\sigma_{tot} \cdot m^2_{WW}}{2\pi} \geq 1$. The ratio $\dfrac{\sigma_{tot} \cdot m^2_{WW}}{2\pi}$ for the events with leading lepton $p_T$ above 760 GeV (940 GeV) for \ohw with $\Lambda=$1000 GeV (1300 GeV) is shown in Fig \ref{eftww}, and we find that the events are well below the unitarity violation threshold. As discussed in\cite{Falkowski:2014tna}, the limits obtained with very high $p_T$ cuts may not be valid since dimension-8 operators may become relevant at these energies. But since we focus on limits from VBF Higgs we use the most optimistic estimate of the limits from WW production, and as discussed in Sec. \ref{vbfsec}, similar limits can be obtained from VBF using much softer cuts thus avoiding the breakdown in validity of the EFT.

\vspace{5mm}

\section{VBF Higgs decaying to photons\label{vbfsec}}

The decay of Higgs to diphotons has a relatively smaller branching ratio; but being a very clean channel the projected sensitivites\cite{ATLASprojection} are similar to the other decay channels. Searches for VBF Higgs typically use a rapidity gap and high dijet mass requirement for the tagging jets to reduce the contribution of Higgs plus vector boson and gluon fusion Higgs plus jets, with the latter still remaining a substantial background. The other backgrounds in the diphoton decay channel are the continuum production of diphotons, and events where jets radiate non-prompt photons known as `fake photons'. Since the selection criteria used by CMS achieves over a 99\% pure source of prompt photons\cite{CMS:rilUpdatedmeasurementsoftheHiggsboson}, we ignore the `fake photon' background. 

The Higgs signal and continuum diphoton background samples were generated using {\tt MadGraph5}, interfaced with {\tt Pythia} and {\tt Delphes}. The background from gluon fusion higgs plus two jets was simulated with {\tt VBFNLO 2.7.0}\cite{vbfnlo,vbfnlomain}, interfaced with {\tt Pythia 8.2}\cite{pythia8} and {\tt Delphes}. For the 7 TeV search with 5.1fb$^{-1}$ of data, we implement the kinematic cuts used by the CMS study\cite{CMS7tevdiphoton} and find that the operator \ohw with $\Lambda=$150 GeV can be excluded at 95\% CL. For the $\sqrt{s}=$ 14TeV analysis we use the following cuts from the updated CMS VBF study\cite{CMS:rilUpdatedmeasurementsoftheHiggsboson}.

Two jets with invariant mass of at least 500 GeV and $p_T>30$ GeV are required, with $\Delta\eta>3$ within $|\eta|<4.7$.  Two photons with $p_T>30$GeV, $|\eta|<2.5$ are required and the azimuthal angle between the diphoton system and dijet system is required to be greater than 2.6. The leading (trailing) photon is required to have $p_T$ greater than $\frac{m_{\gamma\gamma}}{2}$ (25 GeV) with $|\eta|<2.5$. The Zeppenfeld variable $\eta(\gamma_1+\gamma_2)-\frac{\eta_{j_1}+\eta_{j_2}}{2}$ is required to be less than $2.6$. The tagging jets are required to have $\Delta R>0.5$ from the photons.  Taking advantage of the much higher event rates, we impose a cut on the diphoton invariant mass $|m_{\gamma\gamma}-125|<5$GeV. Increasing the dijet mass and jet $p_T$ requirements used in the 8 TeV search did not significantly improve the limits at 14 TeV. The signal and background yields passing these cuts for the SM and a benchmark point are given in Table 2. 

\begin{table}[H]

\centering \begin{tabular}{c|p{3cm}|p{3cm}}
Process & VBF cuts (see text) & $p_T(\gamma\gamma) > 140$ GeV  \\
\hline
SM VBF Higgs$\rightarrow\gamma\gamma$ & 1.27 & 0.408 \\
SM VBF+\ohw $\Lambda =$600 GeV, $c_{HW}=-1$ & 1.35 & 0.534 \\
Continuum diphoton & 3.03 & 0.186   \\
Gluon fusion plus two jets & 0.237 & 0.055 \\
\end{tabular} 
\label{diphotxsec} \caption{VBF Higgs search cross sections (fb) at $\sqrt{s}=14$TeV. Second column shows improved enhancement from \ohw relative to SM with diphoton $p_T$ cut}
\end{table}

\begin{figure}[H]
\centering \includegraphics[scale=0.35]{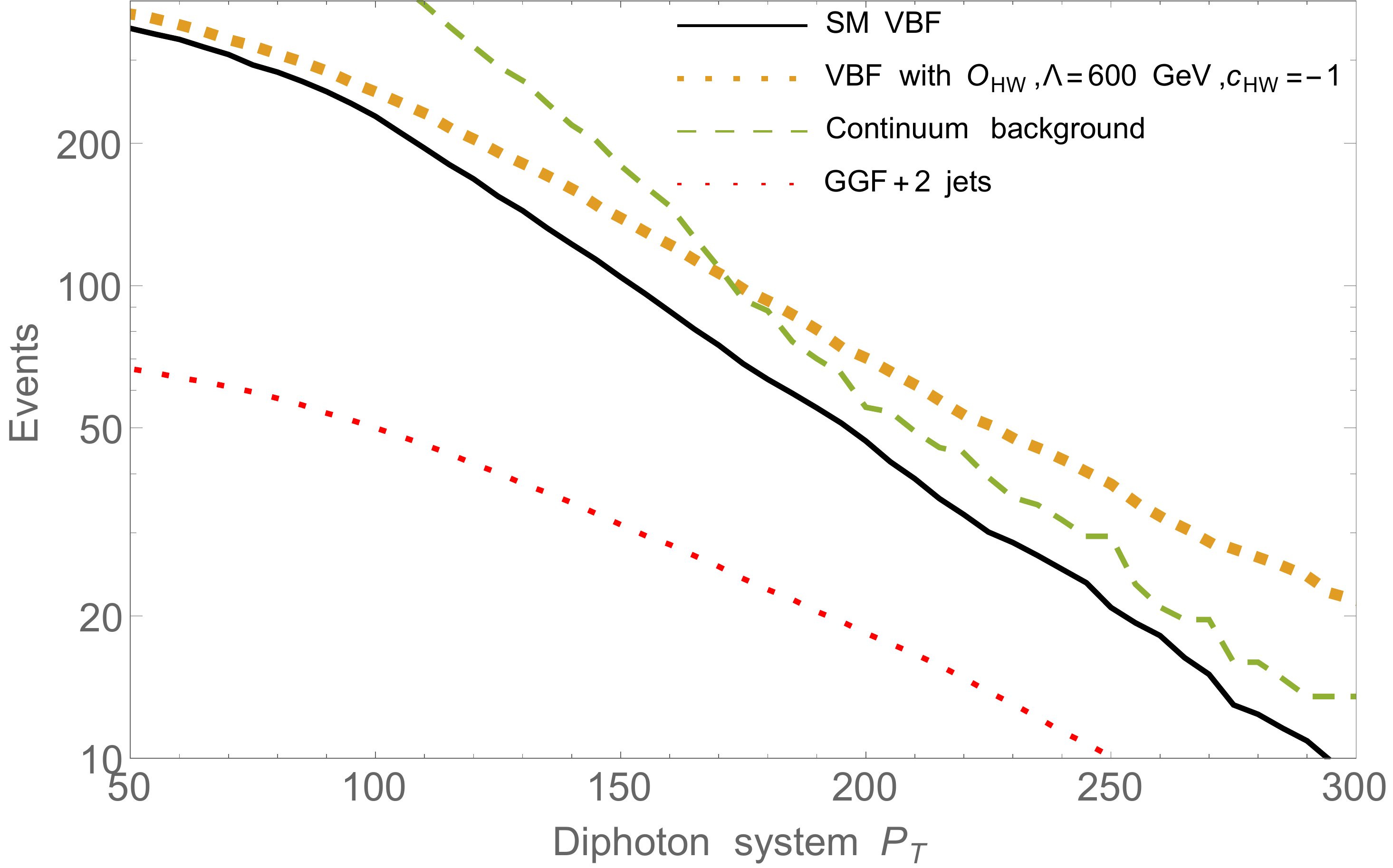}
\begin{minipage}{5.5in}
\caption{\small 
Cumulative plot for total events above given value of $p_T$ of the diphoton system, for $\sqrt{s}=14$TeV and 300fb$^{-1}$}
\label{pltpot}
\end{minipage}
\end{figure}

\begin{figure}
\centering \includegraphics[scale=0.35]{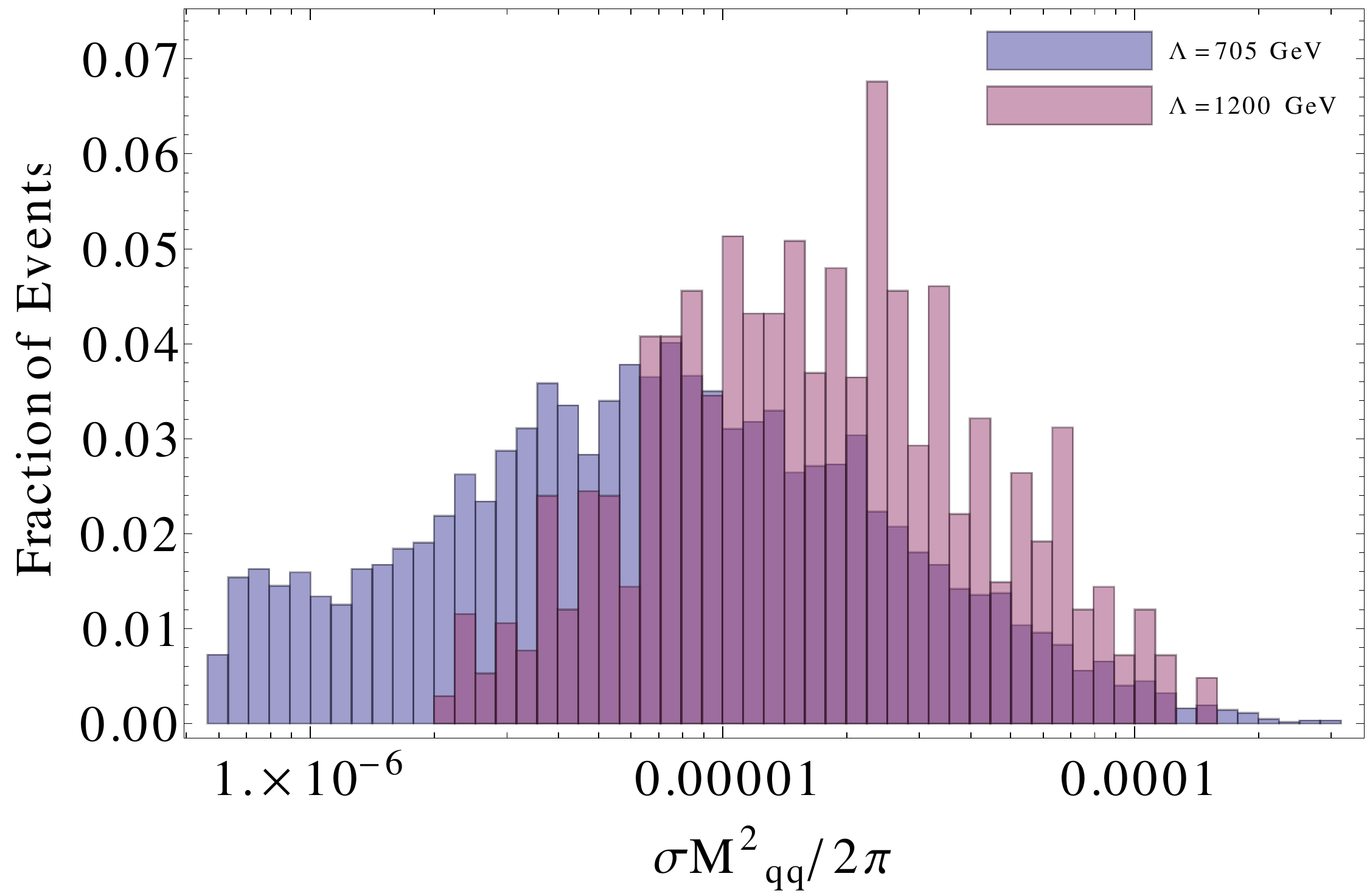}
\caption{\small 
Histogram of $\dfrac{\sigma_{tot} \cdot m^2_{qq}}{2\pi}$ for events with \ohw and $\Lambda=$ 705 GeV (1200 GeV) passing Higgs $p_T$ cut of 140 GeV (330 GeV) respectively}
\label{eftvbf}
\end{figure}

The operator \ohw leads to a new Lorentz structure in the HVV vertices which does not occur in the SM. The vertex with W-bosons, which has the greatest contribution in VBF production is  $$\dfrac{g^2 v c_{HW} }{2\Lambda^2} ( p_1^{\mu} p_2^\nu+p_1^{\mu} p_3^\nu-g^{\mu\nu}(p_1\cdot p_2 +p_1\cdot p_3) ).$$ Due to this momentum dependence the relative enhancement from the operator \ohw is greater at high Higgs $p_T$, as shown in Fig. \ref{pltpot}. Thus the limits can be improved with a cut on the $p_T$ of the Higgs reconstructed from the diphoton system. Using a similar reasoning as in the case of WW production, we have checked that the use of the EFT does not violate unitarity with the given $p_T$ cuts, as shown in Fig. \ref{eftvbf}. The strongest limit that can be obtained in each case with the corresponding $p_T$ cuts are given in Table \ref{vbfxsec}.

\begin{table}[H]
\caption{95\% CL Limits on $\Lambda$ from VBF diphoton search}
\centering \begin{tabular}{c|c|c|c|c}
\label{vbfxsec}

$\sqrt{s}$ and Luminosity &  \multicolumn{2}{c|}{$c_{HW}=-1$} & \multicolumn{2}{c}{$c_{HW}=1$} \\
\hline
& $p_T$ cut & $\Lambda$ & $p_T$ cut & $\Lambda$  \\
\cline{2-5}
7 TeV, 5.1fb$^{-1}$ & no $p_T$ cut & 150 GeV & no $p_T$ cut & 120 GeV \\
14 Tev, 300fb$^{-1}$ & no $p_T$ cut & 510 GeV & no $p_T$ cut & 550 GeV \\
14 Tev, 300fb$^{-1}$ & 140 GeV & 705 GeV & 130 GeV & 600 GeV \\
14 Tev, 3ab$^{-1}$ & 330 GeV & 1200 GeV & 275 GeV & 1200 GeV
\end{tabular}
\end{table}

We note an important feature of the results. The $p_T$ cuts required to maximize the sensitivity to these operators are substantially lower than in the case of WW production. For example, to obtain a 95\% CL limit of $\Lambda > 1200$ GeV (for $c_{HW}=1$) with 3ab$^{-1}$ of data at the 14 TeV LHC, a $p_T$ cut of 670 GeV is required on the leading lepton in WW production, while the same limit can be obtained from VBF with a $p_T$ cut of 330 GeV on the diphoton system.

The constraints on these operators from the signal strength ratio of VBF to gluon fusion $\frac{\mu_{VBF}}{\mu_{ggF}}$ was studied in \cite{Ellis:2014CompleteHiggsSectorConstraintsonDimension6}, using the likelihoods reported by CMS and ATLAS. They obtain a stronger limit on \ohw from 8 TeV data since they combine the signal strengths from $\gamma\gamma,\bar{\tau}\tau,ZZ^*$ and $WW^*$. We have shown that a stronger limit can be obtained in the diphoton channel using a $p_T$ cut, and this should apply to the other channels also.

%Using the UFO implementation  of dim-6 operators from \cite{HELmodel} in {\tt MadGraph5 v2.1.2} interfaced with {\tt Pythia 6.4} and {\tt Delphes 3.1.2} and applying VBF cuts only on the tagging jets with the Higgs decaying to all final states, we obtain Eq. \ref{total} for the signal strength ratio. But if the cuts are applied prior to the Higgs decay at the generator level, we obtain a much stronger enhancement shown in Eq. \ref{gener} and a similar result \ref{diphoteq} is obtained if the cuts are applied to the diphoton final state. This is because the $\bar{b}b$ final state has a much lower efficiency, and we find that an analysis with cuts specific to the final states can yield a stronger limit than generic cuts on only the tagging jets.
%
% \begin{equation}\label{total}
% \dfrac{\sigma(VBF\rightarrow hjj)_{SM+\mathcal{O}_{HW}}}{\sigma(VBF)_{SM}} \simeq 1 - 6.9(\bar{c}_{HW}+tan^2_{\theta_W}\bar{c}_{HB})
% \end{equation}
% \begin{equation}\label{gener}
%\dfrac{\sigma(VBF\rightarrow hjj)_{SM+\mathcal{O}_{HW}}}{\sigma(VBF\rightarrow\gamma\gamma jj)_{SM}} \simeq 1 - 10.9 (\bar{c}_{HW}+tan^2_{\theta_W}\bar{c}_{HB}) 
% \end{equation}
%\begin{equation}\label{diphoteq}
%\dfrac{\sigma(VBF\rightarrow\gamma\gamma jj)_{SM+\mathcal{O}_{HW}}}{\sigma(VBF\rightarrow\gamma\gamma jj)_{SM}} \simeq 1 - 10.57 (\bar{c}_{HW}+tan^2_{\theta_W}\bar{c}_{HB}) 
% \end{equation}

\section{Conclusion}
One of the primary goals of the LHC is to probe deviations of the Higgs from SM expectations, and an effective field theory composed of dimension-6 operators is a particularly useful model-independent framework to parametrize effects from new physics. VBF Higgs measurements are an important probe that is complementary to other methods such as Higgs branching ratio and TGC measurements. As discussed in \cite{Falkowski:2014tna}, in the high $p_T$ tail of WW production, the contribution from dimension-8 operators becomes relevant and hence the limits from WW production derived in \cite{ATLAS:2012Measurementof$W^+W^-$production,Ellis:2014TheEffectiveStandardModelafterLHC} may not be reliable. But VBF searches may avoid this issue, since the $p_T$ cuts required to obtain comparable limits is substantially lower.

The combined limit from the various decay channels of VBF Higgs can be the most sensitive probe of the direction which is allowed by $P_{LR}$ symmetry, $\mathcal{O}_{HW}+\mathcal{O}_{HB}$. We have shown that studying the $p_T$ distribution can substantially increase the sensitivity of VBF measurements to new physics, and Higgs $p_T$ cuts of up to 140 GeV (330 GeV) can maximize the reach of the 14 TeV LHC with 300fb$^{-1}$ (3 ab$^{-1}$) of data. Thus we advocate that CMS and ATLAS optimize and search for dimension-6 operator signals in boosted VBF Higgs. 

\section*{Acknowledgements}   

We thank Markus Luty and Ennio Salvioni for helpful discussions and comments about the manuscript, and Nikolas Kauer for {\tt gg2VV} event samples. This research was supported in part by the Department of Energy under grant DE-FG02-91ER40674

\bibliographystyle{utphys}
\bibliography{vbfhiggs}

\end{document}